\documentclass[aps,prb,twocolumn,a4]{revtex4-1}
\pdfoutput=1
\usepackage{amsmath,amssymb,amsfonts,upgreek}
\usepackage{graphicx}
\usepackage{setspace}
\usepackage{bm}
\usepackage{multirow}

\usepackage{color}
\usepackage{booktabs}

\newcommand{\btfo}{\ensuremath{\rm Bi_5FeTi_3O_{15}}}
\newcommand{\bio}{\ensuremath{\rm Bi_2O_2}}

\newcommand{\fe}{\ensuremath{\rm Fe^{3+}}}
\newcommand{\ti}{\ensuremath{\rm Ti^{4+}}}
\newcommand{\highsym}{\ensuremath{I4/mmm}}
\newcommand{\lowsym}{\ensuremath{A2_1am}}

\graphicspath{{figures/}}

\begin{document}

\title{The potentially multiferroic Aurivillius phase \btfo: cation
  site preference, electric polarization, and magnetic coupling from
  first-principles}

\author{Axiel Ya\"el Birenbaum} 
\email{yael.birenbaum@mat.ethz.ch} 
\affiliation{Materials Theory, ETH Z\"urich, Wolfgang-Pauli-Strasse
  27, 8093 Z\"urich, Switzerland} 
\author{Claude Ederer} 
\email{claude.ederer@mat.ethz.ch} 
\affiliation{Materials Theory, ETH Z\"urich, Wolfgang-Pauli-Strasse
  27, 8093 Z\"urich, Switzerland}

\date{\today}

\begin{abstract}
We study the structural, ferroelectric, and magnetic properties of the
potentially multiferroic Aurivillius phase material \btfo\ using first
principles electronic structure calculations.  Calculations are
performed both with PBE and PBEsol exchange correlation
functionals. We conclude that PBE systematically overestimates the
lattice constants and the magnitude of the ferroelectric distortion,
whereas PBEsol leads to good agreement with available experimental
data. We then assess a potential site preference of the \fe cation by
comparing 10 different distributions of the perovskite $B$-sites. We
find a slight preference for the ``inner'' site, consistent with
recent experimental observations.  We obtain a large value of
$\sim$55\,$\mu$C/cm$^2$ for the spontaneous electric polarization,
which is rather independent of the specific Fe distribution. Finally,
we calculate the strength of the magnetic coupling constants and find
strong antiferromagnetic coupling between \fe cations in nearest
neighbor positions, whereas the coupling between further neighbors is
rather weak. This poses the question whether magnetic long range order
can occur in this system in spite of the low concentration of magnetic
ions.
\end{abstract}

\maketitle

\section{Introduction}
\label{sec:intro}

Multiferroic materials, which exhibit simultaneous ferroelectric and
magnetic order, have been attracting considerable attention during the
last decade.\cite{Spaldin/Fiebig:2005,Eerenstein/Mathur/Scott:2006,Cheong/Mostovoy:2007,Ramesh/Spaldin:2007,Spaldin/Cheong/Ramesh:2010}
Much of the corresponding research is motivated by the great potential
of these materials for non-volatile data storage, sensor applications,
and many other areas of technology. However, materials that exhibit
multiferroic properties above room temperature, and are thus suitable
candidates for future device applications, are still extremely
scarce. The search for alternative materials with robust multiferroic
properties above room temperature is therefore of high relevance.

\begin{figure}
\includegraphics[width=0.25\columnwidth]{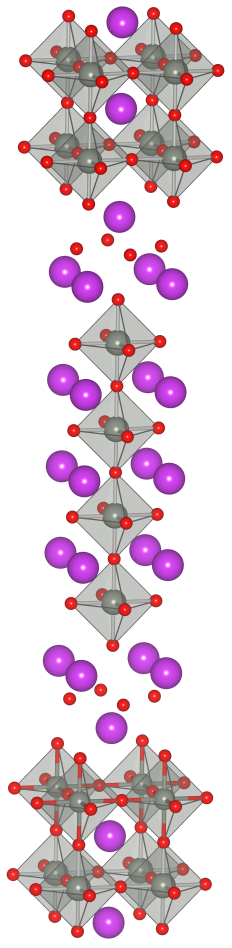}
\caption{(Color online). High symmetry tetragonal crystal structure of
  \btfo\,(space group \highsym), corresponding to $m=4$ in the
  Aurivillius series. Four perovskite-like layers are stacked along
  [001] and separated by fluorite-like $ \rm \left(Bi_2O_2
  \right)^{2-}$ layers. Bi and O atoms are shown as large and small
  (purple and red) spheres, respectively, whereas the Fe/Ti atoms are
  located inside the gray octahedra. Note that each perovskite block
  is offset relative to the next one by $\left(\frac{a_0}{2},
  \frac{a_0}{2}, \frac{c_0}{2}\right)$ where $a_0$ and $c_0$ are the
  lattice parameters of the conventional tetragonal unit cell. This
  picture has been generated using VESTA.\cite{Momma/Izumi:2011}}
\label{tetragonal_Aurivillius}
\end{figure}

One possible route for the design of novel multiferroic materials is
to start from a series of well-established ferroelectrics and create
additional functionality by incorporating magnetic ions into these
systems. A  promising class of materials for this purpose are the
so-called \emph{Aurivillius phases}. The Aurivillius phases are a
family of compounds that form in a naturally-layered
perovskite-related crystal structure which consists of $m$ perovskite
layers $\left(A_{m-1}B_m{\rm O}_{3m+1}\right)^{2-}$, stacked along the
$[001]$ direction, and separated by fluorite-like $\rm \left(Bi_2
O_2\right)^{2+}$ layers (see Fig.~\ref{tetragonal_Aurivillius} for an
example with $m=4$). The overall chemical composition is thus
Bi$_2A_{m-1}B_{m}$O$_{3m+3}$, where many different cations can be
incorporated on the $A$ and $B$-sites within the perovskite-like
layers.\cite{Newnham/Wolfe/Dorrian:1971} The Aurivillius phases are
well known for their excellent ferroelectric properties with very low
fatigue,\cite{PazdeAraujo_et_al:1995,Park_et_al:1999} and offer great
potential for tailoring specific properties by varying both ionic
composition and number of layers.

A number of Aurivillius compounds exhibiting multiferroic properties
have been reported recently. For example, room temperature
ferromagnetism together with ferroelectric polarization has been
reported for materials based on the 4-layer compound \btfo, but with
half of the Fe cations substituted by either Co or
Ni.\cite{Mao_et_al:2009,Chen_et_al:2014} Furthermore, magnetic
field-induced ferroelectric domain switching has recently been
observed in a 5-layer system with composition
Bi$_6$Ti$_{2.8}$Fe$_{1.52}$Mn$_{0.68}$O$_{18}$.\cite{Keeney_et_al:2013}
However, it should be noted that in many cases the observed
polarization and magnetization depend strongly on the synthesis method
and annealing conditions,\cite{Bai_et_al:2012,Mao_et_al_JMS:2012} and
it has been pointed out that the weak magnetic signals found in these
samples can indeed be caused by tiny inclusions of other phases, which
are extremely hard to detect using standard characterization
methods.\cite{Keeney_et_al:2012}

Even in the case of \btfo, the ``parent compound'' for many of the
compositions that have been explored as potential multiferroics, the
ferroelectric and magnetic properties are not well established. An
antiferromagnetic N{\'e}el temperature of 80\,K has been
reported.\cite{Srinivas_et_al:1999} However, several more recent
studies show paramagnetic behavior with no magnetic long-range order
even at very low
temperatures.\cite{Dong_et_al:2008,Bai_et_al:2012,Jartych_et_al:2013,Chen_et_al:2013}
Fits of magnetic susceptibility data have found conflicting signs of
the Curie-Weiss temperature in different samples, indicating either
ferromagnetic\cite{Bai_et_al:2012} or
antiferromagnetic\cite{Dong_et_al:2008,Chen_et_al:2013} interactions
between the Fe cations. Similarly, no well-established value exists
for the spontaneous electric polarization in \btfo, due to
difficulties in obtaining fully saturated ferroelectric hysteresis
loops. Reported values for the remanent polarization vary from $\rm
3.5~\mu C/cm^2$ to $\rm \sim 30~\mu
C/cm^2$.\cite{Mao_et_al_JMS:2012,Bai_et_al:2012}

It is therefore  desirable to establish the intrinsic properties
of \btfo\, and other potentially multiferroic Aurivillius phases using
modern first-principles electronic structure calculations. Such
calculations provide a useful reference for experimental observations
and can also be used as guideline for future studies.

Here we present results of such first-principles calculations for the
4-layer Aurivillius phase \btfo. We show that the structural
properties of this material can be obtained in good agreement with
available experimental data using the PBEsol functional for the
exchange-correlation energy. We also show that there is a preference
for the \fe\, cation to occupy the ``inner'' sites within the
perovskite-like layers, and we suggest the possibility to reverse this
preference using epitaxial strain. Furthermore, we calculate a large
spontaneous ferroelectric polarization of $\sim$55\,$\mu$C/cm$^2$,
which is similar in magnitude to the related 3-layer system
Bi$_4$Ti$_3$O$_{12}$. This shows that the presence of the magnetic
\fe\, cations is not detrimental to the ferroelectric properties of
\btfo. Finally, we calculate the strength of the magnetic coupling and
find a very strong and antiferromagnetic coupling between Fe in
nearest-neighbor positions. However, the coupling becomes nearly
negligible beyond second neighbors, consistent with the short range
character of the underlying superexchange mechanism. This leaves the
question of whether long-range magnetic order can be expected in
\btfo\, open for future studies.

This paper is organized as follows. In the next section we present the
method we use in our first-principles calculations, and discuss in
particular how we treat the quasi-random distribution of \fe\, and
\ti\, cations over the available sites within the perovskite-like
layers. In Sections~\ref{subsec:global_structure} and
\ref{subsec:internal_structure} we then present our results for the
structural properties and the energetics of the different cation
distributions. Sec.~\ref{subsec:pol} focuses on the calculation of the
electric polarization, and in Sec.~\ref{subsec:dos} we give a brief
discussion of the electronic structure in \btfo. Finally, in
Sec.~\ref{subsec:magnetism} we present our analysis of the calculated
magnetic coupling constants and discuss the possibility of long-range
magnetic order. A summary and some further conclusions are given in
Sec.~\ref{sec:summary}.

\section{Methods}
\label{sec:method}

\subsection{Aurivillius crystal structure and different distributions of Fe/Ti cations}
\label{subsec:method_structure}

Most members of the homologous series of Aurivillius compounds exhibit
a tetragonal structure with space group symmetry \highsym\ at high
temperatures (see Fig.~\ref{tetragonal_Aurivillius} for the case of
$m=4$). On cooling, this structure transforms into a polar structure
corresponding to a subgroup symmetry of \highsym.  Even-layered
Aurivillius phases such as \btfo\, (i.e., $m = 2, 4, \ldots$)
transform to orthorhombic symmetry \lowsym\ (space group no. 36),
whereas odd-layered members ($m = 1, 3, \ldots$) transform into $B2cb$
symmetry (space group no. 41) or an even lower symmetry corresponding
to a subgroup of $B2cb$. \cite{Newnham/Wolfe/Dorrian:1971,
  Perez-Mato_et_al:2008} The symmetry lowering from \highsym\ to
\lowsym\ leads to a doubling of the primitive unit cell, with 2
formula units (48 atoms) per unit cell in the \lowsym\ structure.

\begin{table*}
\caption{\label{config}(Color online). The 10 symmetrically inequivalent
  configurations considered in this work, corresponding to different
  distributions of Fe and Ti over the octahedral $B$-sites. Fe and Ti
  sites are indicated by dark (brown) and light (blue) octahedra,
  respectively. The configurations are grouped into ``outer'',
  ``mixed'', and ``inner'', depending on whether two, one, or none of
  the two Fe cations in the unit cell are situated in the perovskite
  layers adjacent to the Bi$_2$O$_2$ layers. The notation defined in
  the second line is used throughout this article to denote the
  various configurations.}
\begin{ruledtabular}
\begin{tabular}{ccc|cccc|ccc}
\multicolumn{3}{c|}{``outer''} 
& 
\multicolumn{4}{c|}{``mixed''} 
& 
\multicolumn{3}{c}{``inner''} 
\\
O1 & O2 & O3 & M1 & M2 & M3 & M4 & I1 & I2 & I3 \\ 
\includegraphics[width=25pt]{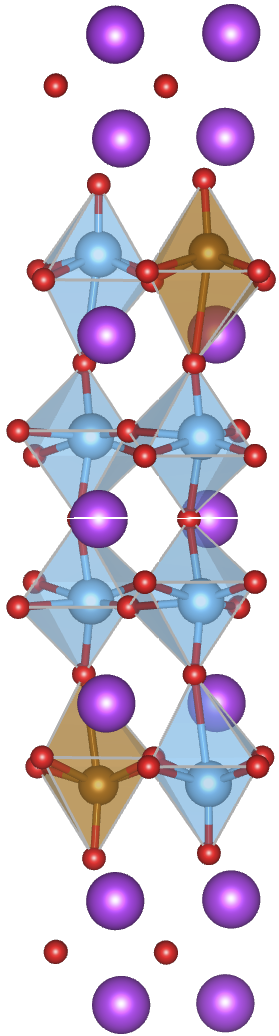}
& 
\includegraphics[width=25pt]{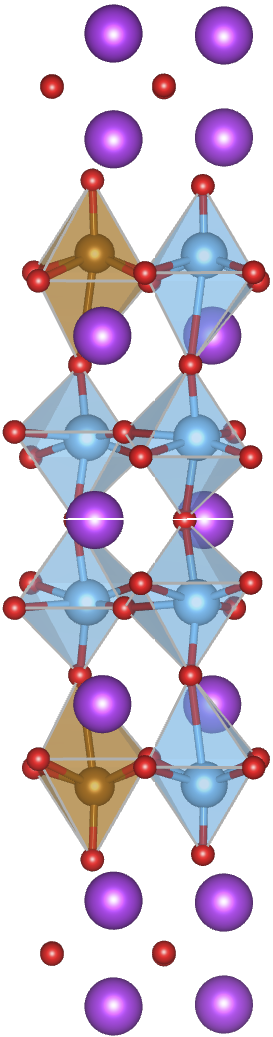}
& 
\includegraphics[width=25pt]{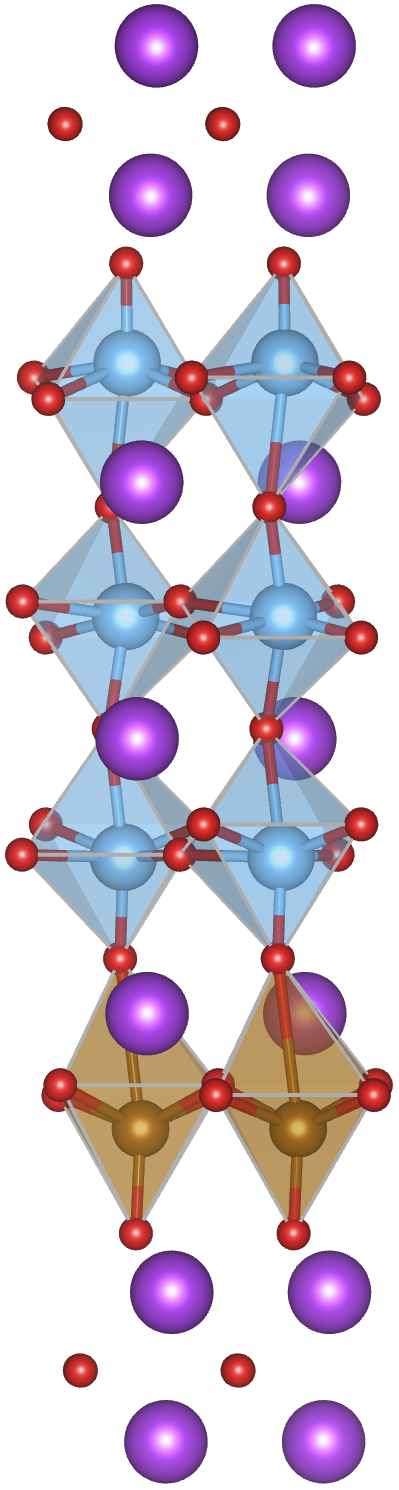}
&
\includegraphics[width=25pt]{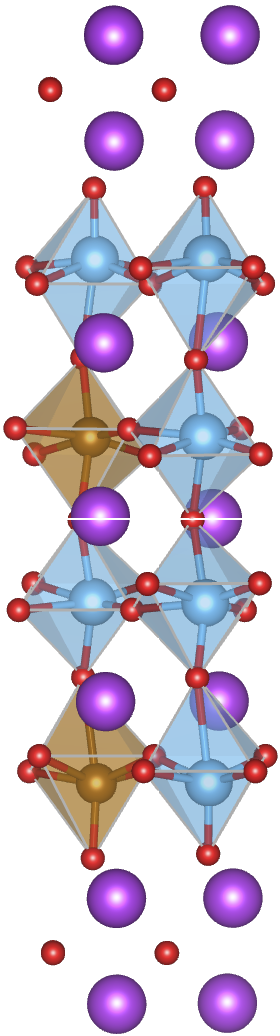}
&
\includegraphics[width=25pt]{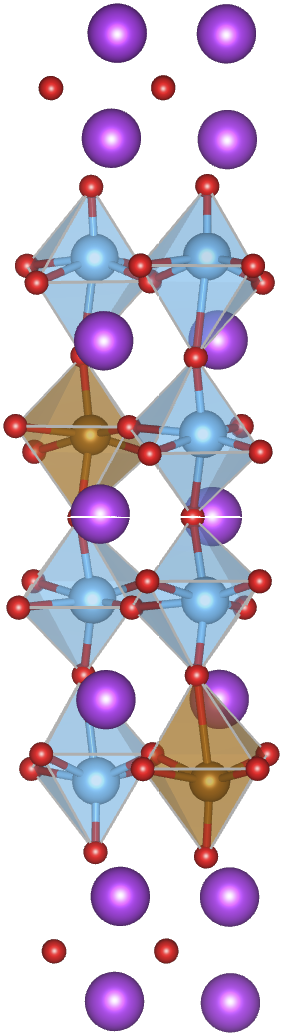}
&
\includegraphics[width=25pt]{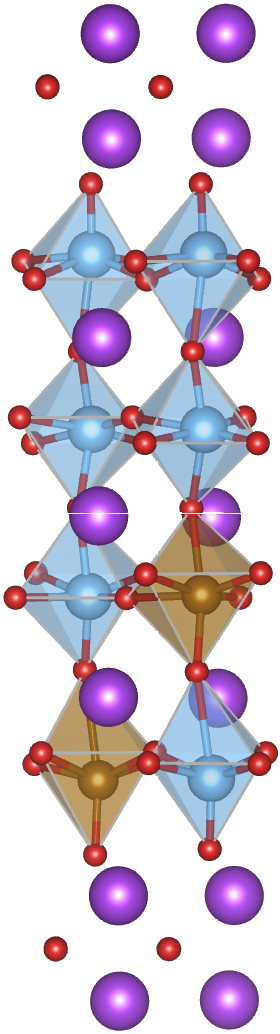}
&
\includegraphics[width=25pt]{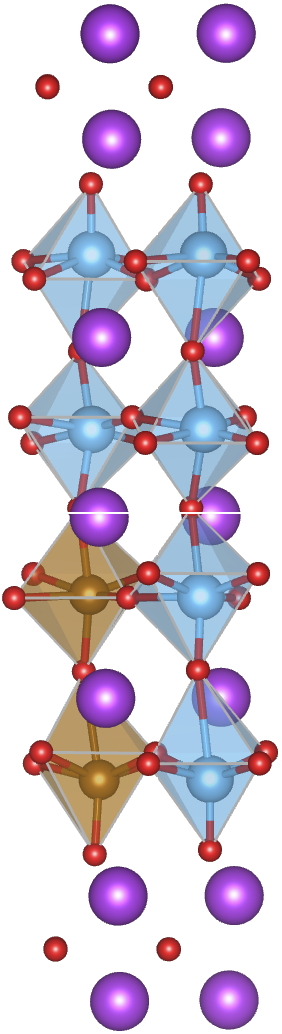}
&
\includegraphics[width=25pt]{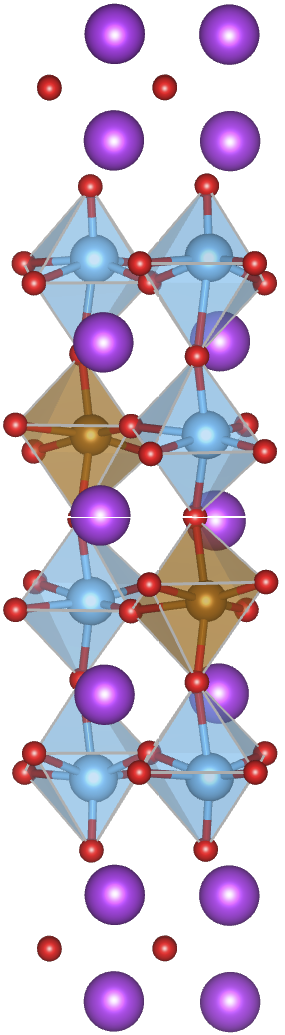}
&
\includegraphics[width=25pt]{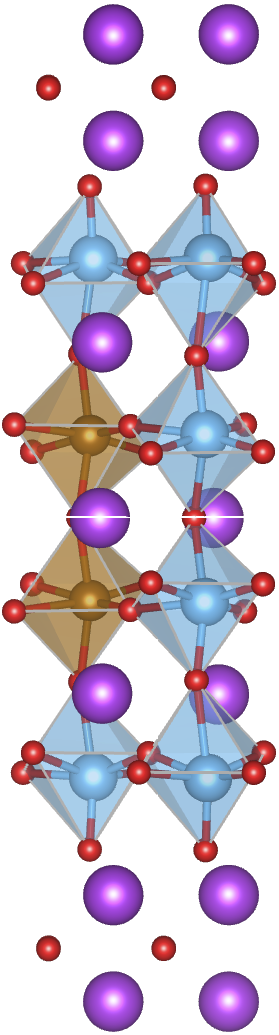}
&
\includegraphics[width=25pt]{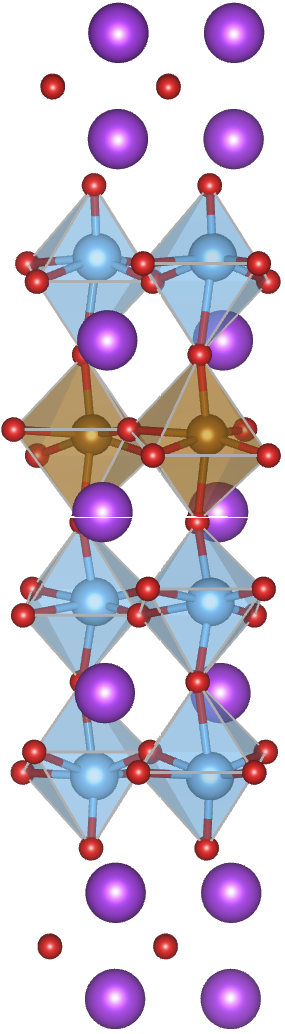} 
\\
\end{tabular}
\end{ruledtabular}
\end{table*}

There are two inequivalent $B$-sites in the 4-layer structure: the
\emph{inner} sites within the perovskite layers, and the \emph{outer}
sites adjacent to the Bi$_2$O$_2$ layers. In \btfo\ these octahedrally
coordinated $B$-sites are shared between the \fe\ and \ti\ cations,
and a quasi-random cation distribution is observed in experiments.
\cite{Hervoches_et_al:2002} However, recent M\"ossbauer experiments
have reported a slight preference of \fe\ to occupy the inner sites
within the perovskite layers, whereas the outer sites are
preferentially occupied by \ti\ cations.\cite{Lomanova_et_al:2012}

In our calculations for \btfo\ we use the 48 atom unit cell
corresponding to the primitive unit cell of the low-symmetry
\lowsym\ structure, which contains 8 octahedrally-coordinated $B$
sites (4 inner and 4 outer sites). There are 10 symmetrically
inequivalent ways to distribute 2 Fe and 6 Ti atoms over these 8 $B$
sites, all of which are depicted in Table~\ref{config}. In this work
we explicitly consider each of the 10 individual configurations listed
in Table~\ref{config}. This allows us to evaluate the impact of a
specific cation distribution on the physical properties and also to
address a potential site preference of the \fe\ and \ti\ cations in
\btfo. We note that the specific distribution of cations breaks
further symmetries compared to the experimental
\lowsym\ structure. One could think of the latter as a weighted
average over all the possible configurations.

To allow for all distortions observed in the experimental
\lowsym\ symmetry, we use lattice vectors that resemble the
corresponding base-centered orthorhombic lattice vectors. In other
words, $\vec{a} \approx (\sqrt{2}a_0, 0, 0)$, $\vec{b} \approx (0,
\sqrt{2}a_0, 0)$ and $\vec{c} \approx (\sqrt{2}a_0/2, 0, c_0/2)$,
where $a_0$ and $c_0$ correspond to the in-plane and out-of-plane
lattice constants of the high-symmetry
\highsym\ structure.\footnote{Note that strictly speaking the space
  group notation \lowsym\ implies a setting where the polar screw axis
  $2_1$ is along the $\vec{a}$ direction, whereas in our calculations
  we use a setting with the polar axis along $\vec{b}$. For
  consistency with the literature we nevertheless use \lowsym\ to
  denote space group no. 36 within this article.} Due to the lower
symmetry of the individual cation configurations listed in
Table~\ref{config}, we obtain small deviations from the ideal
base-centered orthorhombic lattice after relaxation. Specifically, the
angles between the lattice vectors deviate slightly from the angles
within perfect orthorhombic symmetry. However, in all cases these
deviations are smaller than 0.7$^\circ$.

\subsection{Computational details}
\label{subsec:method_details}

We perform first-principles calculations using density functional
theory (DFT) and the projector augmented wave (PAW) method as
implemented in the Vienna \textit{ab initio} simulation package
(VASP).\cite{Kresse/Furthmueller_CMS:1996,Kresse/Joubert:1999} We use
the generalized gradient approximation according to Perdew, Burke, and
Ernzerhof (PBE) and its version optimized for solids
(PBEsol).\cite{Perdew/Burke/Ernzerhof:1996,Perdew_et_al:2008} We use
PAW potentials which include 15 valence electrons for Bi ($6s^2
5d^{10} 6p^3$), 14 for Fe ($3p^6 4s^2 3d^6$), 10 for Ti ($3p^6 4s^2
3d^2$), and 6 for O ($2s^2 2p^4$). To correctly treat the strong
interactions between the Fe $d$ electrons, the Hubbard ``$+U$''
correction was applied with $U_\text{Fe} = 3.0$ eV and the Hund's
coupling parameter $J_H$ set to
zero.\cite{Anisimov/Aryatesiawan/Liechtenstein:1997} This value for
$U_\text{Fe}$ lies in the typical range that has been used
successfully to describe various properties of
BiFeO$_3$.\cite{Neaton_et_al:2005, Kornev_et_al:2007,
  Hatt/Spaldin/Ederer:2010, Dieguez_et_al:2011} Some calculations with
$U_\text{Fe} = 5$\,eV are also performed to assess the sensitivity of
electronic structure and magnetic coupling constants on the magnitude
of $U_\text{Fe}$.

We perform full structural relaxations of all degrees of freedom
(lattice parameters and internal positions) for all 10 configurations
listed in Table~\ref{config}, both with parallel and antiparallel
orientation of the magnetic moments of the two \fe\, cations within
the unit cell. Lattice parameters and ionic positions are relaxed
until the residual forces are smaller than $10^{-3}$ eV/\AA\ and the
total energy changes less than $10^{-8}$ eV. Calculations are
converged using a $\Gamma$-centered $k$-point mesh with $4 \times 4
\times 2$ divisions along the three reciprocal lattice vectors and a
plane wave cutoff energy of $E_{\rm cut} = 550$\,eV. For the
calculation of the electronic density of states a denser (and more
uniform) $12 \times 12 \times 3$ k-point mesh is used.

\section{Results and discussion}
\label{sec:results}

\subsection{Lattice parameters and energetics}
\label{subsec:global_structure}

In this section we present the energies and lattice parameters
obtained by full structural relaxation of all 10 symmetrically
inequivalent configurations with different distributions of \fe\, and
\ti\, cations listed in Table~\ref{config}. All presented results are
obtained for antiparallel alignment of the Fe magnetic moments. As
will be shown in Sec.~\ref{subsec:magnetism} this corresponds to the
preferred magnetic state in all cases.

\begin{figure}
\includegraphics[width=\columnwidth]{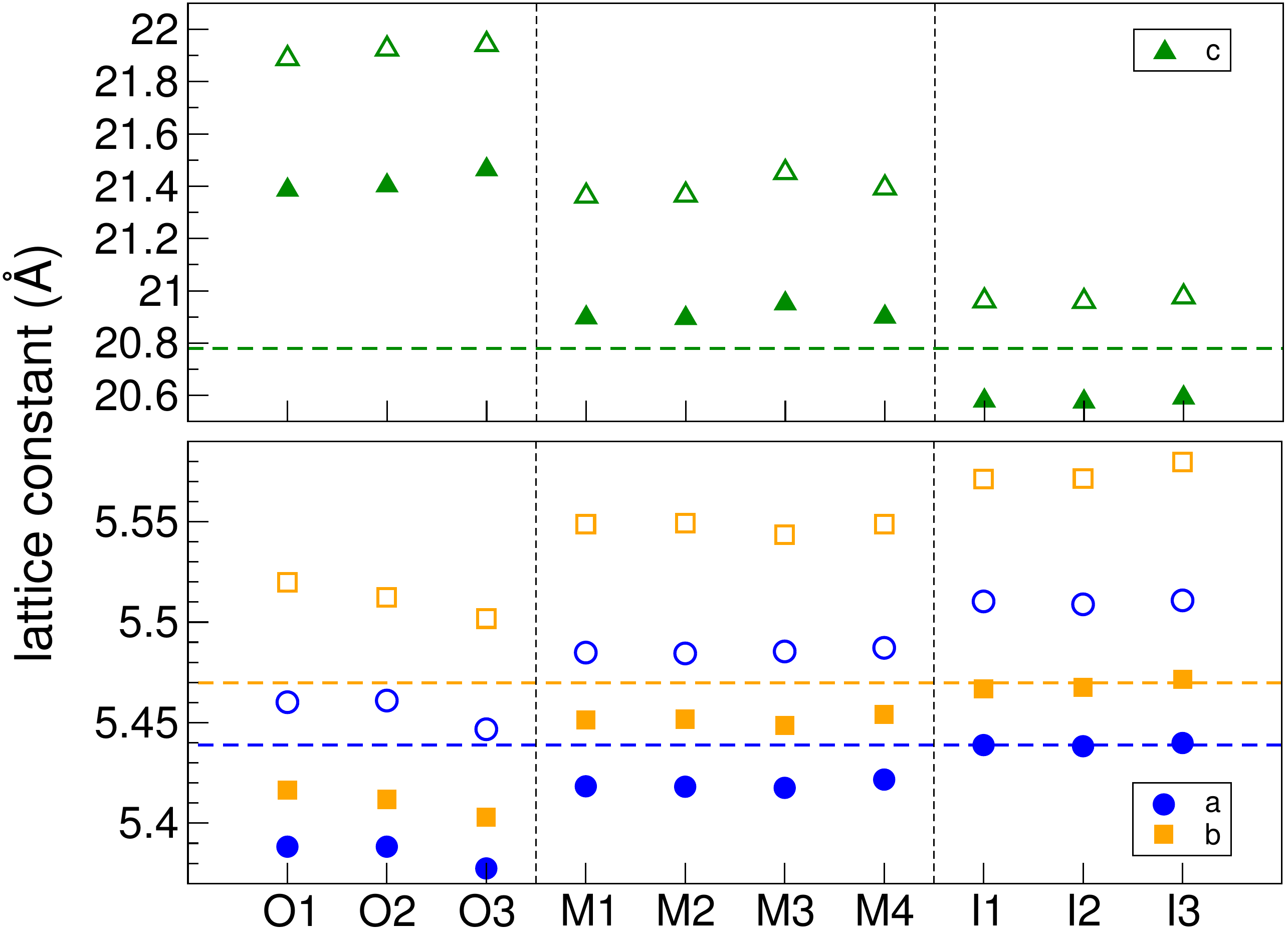}
\caption{(Color online). Lengths of the three lattice vectors
  $a=|\vec{a}|$ (bottom, circles), $b=|\vec{b}|$ (bottom, squares),
  and $c=|\vec{c}|$ (top, triangles) for the 10 different
  configurations listed in Table~\ref{config}, computed using the
  PBEsol (filled symbols) and the PBE (open symbols) functionals. The
  horizontal dashed lines correspond to experimental data from
  Ref.~\onlinecite{Hervoches_et_al:2002}.}
\label{latt-cst}
\end{figure} 

Fig.~\ref{latt-cst} shows the lengths of the three lattice vectors
obtained after full structural relaxation for each configuration, in
comparison with experimental data taken from
Ref.~\onlinecite{Hervoches_et_al:2002}.  One can see that for both PBE
and PBEsol there is a clear correlation between the lattice constants
and the distribution of \fe\, and \ti\, cations over the available $B$
sites.  The three inner configurations have longer in-plane lattice
constants, $a$ and $b$, and shorter out-of-plane constants, $c$,
compared to the three outer configurations. Mixed configurations are
in between these two cases. Differences between configurations of the
same ``type'' (i.e. inner, outer, or mixed) are rather small.

The larger tetragonality (larger $c/a$) of the crystallographic unit
cell for the outer configurations is related to a consistently large
\emph{local} tetragonality of the Fe sites in the outer perovskite
layer. For the following we define the local tetragonality of the
perovskite sites as the ratio between the out-of-plane and in-plane
O-O distances, $d_{\perp}/d_{\parallel}$, of the surrounding oxygen
octahedron. For the outer Fe sites, averaged over the three outer
configurations (O1-O3), we obtain $d_{\perp}/d_{\parallel} = 1.21$
within PBE and $d_{\perp}/d_{\parallel} = 1.16$ within PBEsol. In
contrast, the average outer site tetragonality with all Fe occupying
the inner site (I1-I3) is only $d_{\perp}/d_{\parallel} = 1.04$ with
either exchange correlation functional. Furthermore, in all outer and
mixed configurations the local tetragonality of the outer Fe sites is
consistently larger than that of the outer Ti sites. The local
tetragonality of the inner perovskite sites is always close to 1
(average $d_{\perp}/d_{\parallel} = 1.01$ within PBEsol).

It is interesting to note that the large tetragonal distortion of the
oxygen octahedron together with the large off-center displacement of
the \fe\, cation on the outer site leads to a quasi five-fold
square-pyramid coordination of \fe\, that closely resembles the
``super-tetragonal'' phase found in thin films of BiFeO$_3$ under
strong compressive strain.\cite{Zeches_et_al:2009, Bea_et_al:2009,
  Hatt/Spaldin/Ederer:2010}

While both PBE and PBEsol exhibit the same qualitative trends, the
level of agreement with the experimental lattice parameters is
different for the two functionals. Generally, PBEsol leads to slightly
smaller lattice parameters compared to PBE, and overall seems to
provide better agreement with the experimental data. The best
agreement is observed for the inner and mixed configurations using
PBEsol. On average (over all configurations), PBE overestimates $a$,
$b$, and $c$ by 0.05\,\AA\ (0.8\,\%), 0.08\,\AA\ (1.4\,\%), and
0.63\,\AA\ (3.1\,\%), respectively, whereas PBEsol underestimates both
$a$ and $b$ by 0.02\,\AA\ (0.4\,\%), and overestimates $c$ by
0.18\,\AA\ (0.8\,\%). 

We also verified that the specific value of the Hubbard $U$ does not
have a noticeable effect on the structural properties. Within PBEsol,
an increase of the Hubbard $U$ applied on the Fe site from $U_{\rm Fe}
= 3$\,eV to $U_{\rm Fe} = 5$\,eV leads to differences in lattice
constants and volume of less than $0.1$\%.

\begin{figure}
\includegraphics[width=\columnwidth]{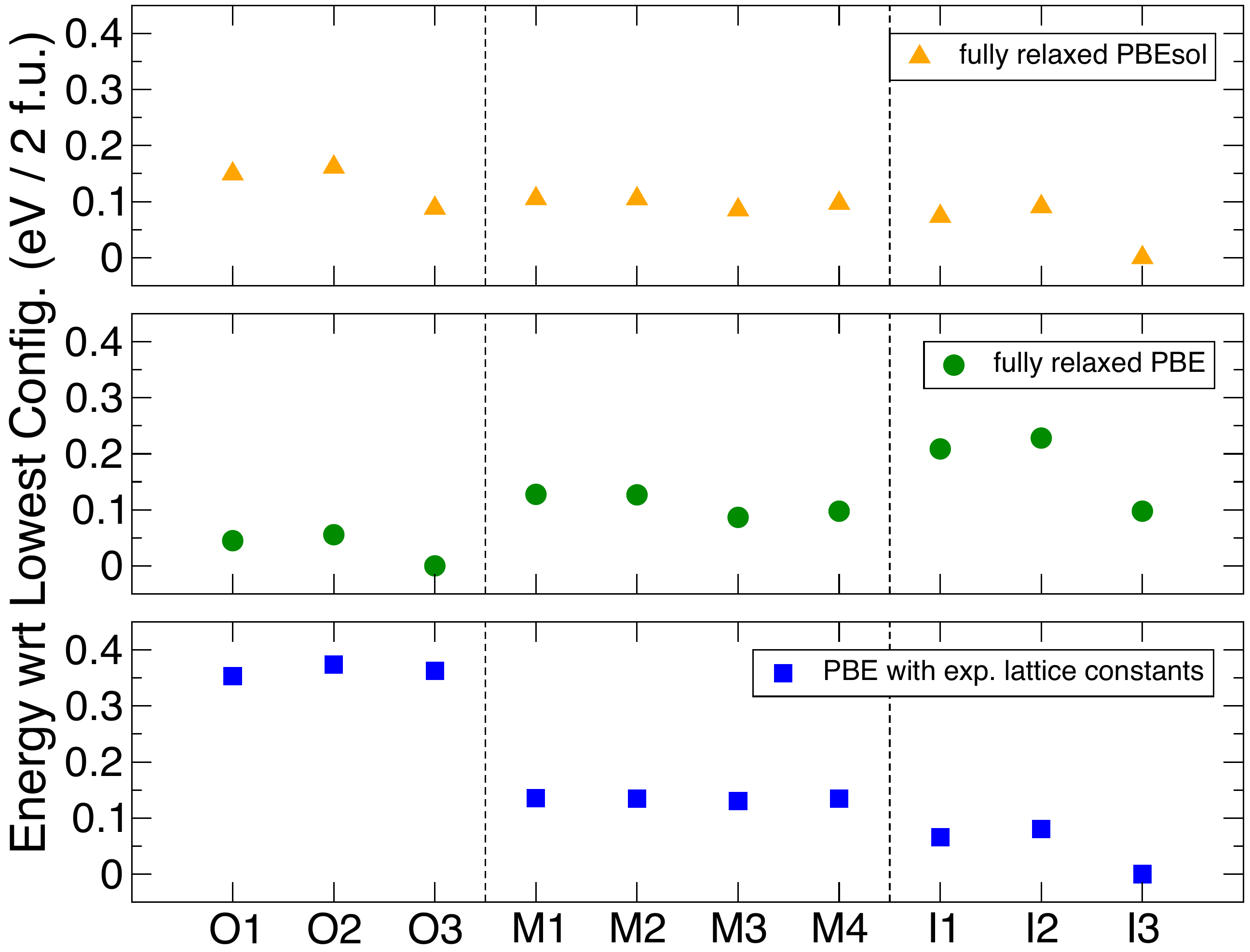}
\caption{(Color online). Relative energies of all configurations for
  the three cases discussed in the main text: full relaxation using
  PBEsol (top), full relaxation using PBE (middle), and only ionic
  positions relaxed using PBE with lattice parameters fixed to
  experimental values (bottom).}
\label{energetics}
\end{figure}

In order to identify a possible site preference of the \fe\ cation, we
next compare the relative energies per unit cell for all
configurations. To estimate the effect of the slight under- or
overestimation of the lattice parameters compared to the experimental
values, we consider three distinct cases: i) full structural
relaxation using the PBEsol functional, ii) full relaxation using PBE,
and iii) lattice parameters fixed to the experimental values from
Ref.~\onlinecite{Hervoches_et_al:2002} and internal atomic coordinates
relaxed using PBE.  We find that in all three cases, the energies of
the different configurations form a hierarchy according to the
classification of the Fe distribution as inner, outer, and mixed (see
Fig.~\ref{energetics}). This indicates that the main factor
determining the energetics is the total amount of Fe sitting in the
inner (or outer) perovskite layers and that the specific spatial
distribution within each layer is less important.  It can further be
seen that the full PBEsol relaxations, which give the best agreement
with the experimental lattice constants, indicate a preference of the
\fe\, cations to occupy the inner site, whereas the full PBE
relaxations, which slightly overestimate the lattice constants, prefer
Fe on the outer sites.  Interestingly, the calculations where the
lattice parameters have been fixed to experimental values and only the
internal positions of the ions are relaxed using PBE gives the same
site preference as PBEsol, i.e. lower energies for the inner
configurations. Therefore, the slight overestimation of the lattice
parameters within PBE seems to be responsible for the different site
preference obtained within PBE and PBEsol.  This sensitivity of the
observed site preference on the lattice parameters as well as the more
elongated unit cell (larger $c/a$ ratio) obtained for the outer
configurations (see Fig.~\ref{latt-cst}) suggest a potential way of
tailoring the cation distribution using epitaxial strain, i.e. by
growing thin films of \btfo\, on different substrates with either
positive or negative lattice mismatch.

As can be seen from Fig.~\ref{energetics}, the energy differences
between different cation configurations are of the order of
$\sim$100\,meV per unit cell. To obtain a rough estimate of whether these
energy differences can lead to noticeable deviations from a random
distribution of \fe\, and \ti\, cations over the octahedral sites, we
calculate the Fe occupation of the outer site as a thermal average,
\begin{equation}
\langle c^{(\text{out})}_\text{Fe}\rangle (T) = \frac{1}{Z} \sum_i
c^{(\text{out})}_{\text{Fe},i} e^{-E_i/k_BT} \quad ,
\end{equation}
where $E_i$ and $c^{(\text{out})}_{\text{Fe},i}$ are the energy and
outer site occupation of configuration $i$, and $Z=\sum_i
e^{-E_i/k_BT}$.

We note that this is a  simplified model that considers
only the 10 configurations that correspond to our choice of periodic
unit cell and does not take into account any kinetic effects during
growth. We therefore do not expect it to accurately predict
the outer site occupation in realistic samples. Nevertheless, this
model gives a rough idea of the temperature range in which a
preferential site occupation can be expected based on simple
energetics.

\begin{figure}
\includegraphics[width=\columnwidth]{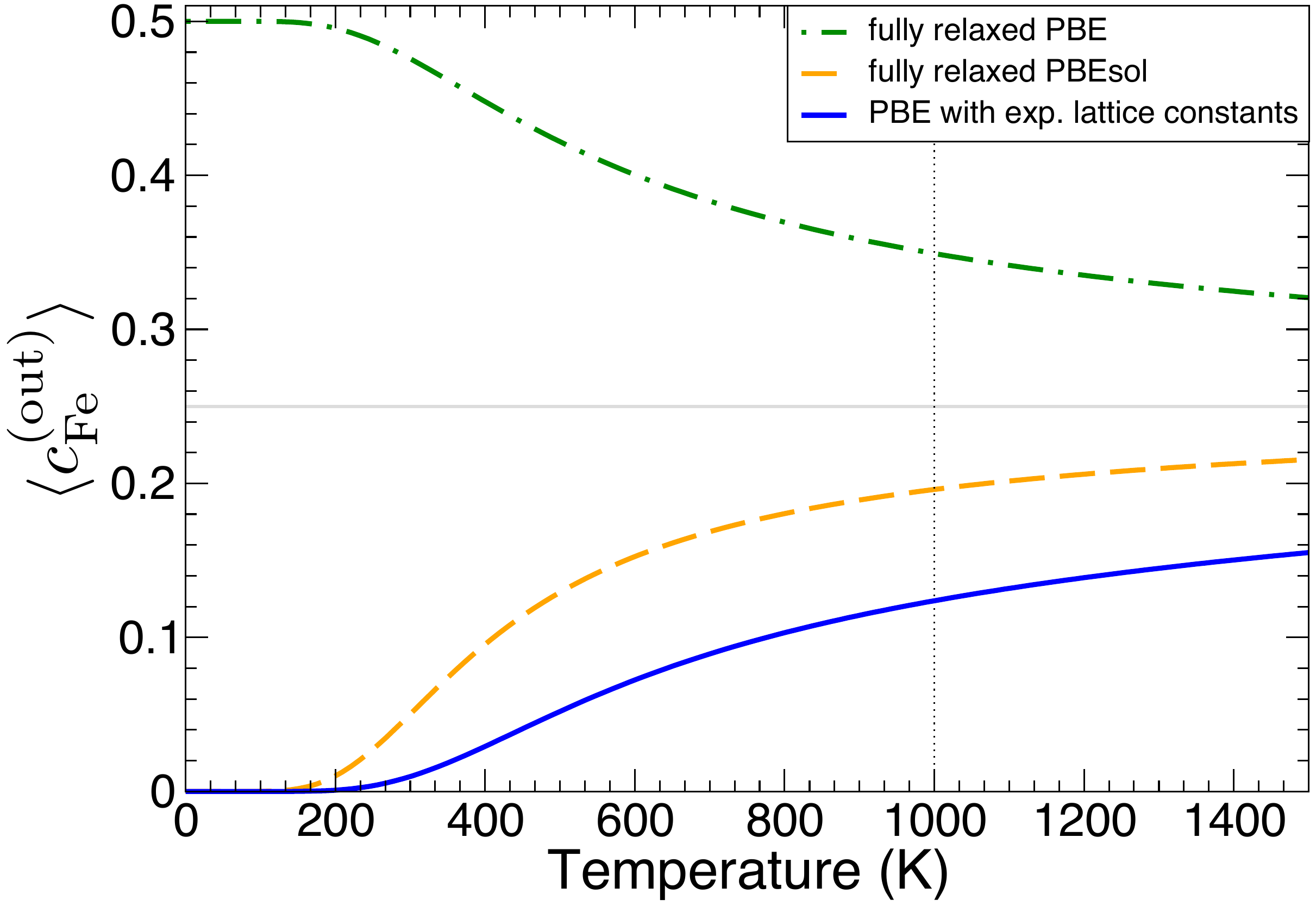}
\caption{(Color online) Thermal average of the Fe occupation on the
  outer site for the three different cases discussed in the text. The
  vertical dotted line indicates a typical growth temperature of
  1000\,K. The horizontal dashed line indicates the value of 0.25
  corresponding to a completely random cation distribution.}
\label{fig:proba_out}
\end{figure}

Fig.~\ref{fig:proba_out} shows the resulting temperature dependence of
$\langle c^{(\text{out})}_\text{Fe}\rangle$, i.e. the average Fe
occupation of the outer site. At zero temperature the value of
$\langle c^{(\text{out})}_\text{Fe} \rangle$ is determined by the
lowest energy configuration in each case (compare with
Fig.~\ref{energetics}), i.e. zero Fe occupation of the outer site
within PBEsol and within PBE using experimental lattice parameters,
while $\langle c^{(\text{out})}_\text{Fe} \rangle = 0.5$ for the fully
relaxed PBE case, corresponding to all Fe cations sitting on an outer
site. Furthermore, for $T \rightarrow \infty$ all three cases converge
to a value of 0.25, which corresponds to a completely random
distribution of \fe\, cations over outer and inner sites. However, it
can be seen that within the given approximations a fully random site
occupancy is not reached even at temperatures significantly above
1000\,K. Indeed, at a typical synthesis temperature of 1000\,K (see
e.g. Refs.~\onlinecite{Hervoches_et_al:2002},
\onlinecite{Bai_et_al:2012}, or \onlinecite{Nakashima_et_al:2010})
less than 20\,\% of outer sites are occupied by \fe\ within PBEsol and
when using experimental lattice parameters. This means that more than
60\,\% of all \fe\ cations can be found on the inner sites. Thus, our
results suggest that it is possible to alter the occupation of outer
and inner sites by varying growth or sintering temperatures, cooling
rates, and annealing times during synthesis.

Experimentally, while no indications for a deviation from the random
distribution of \fe\, and \ti\, cations have been found in Rietveld
refinements of neutron diffraction data,\cite{Hervoches_et_al:2002}
recent M{\"o}ssbauer spectroscopy indicates a small preference of the
\fe\, cations for the inner perovskite
sites,\cite{Lomanova_et_al:2012} consistent with our PBEsol
results. Lomanova \emph{et al.} also found that this preferential
occupation of inner (outer) sites with \fe\ (\ti) strengthens as the
number of perovskite-like layers increases ($m \geqslant 5$) within
the series
Bi$_{n+1}$Fe$_{n-3}$Ti$_3$O$_{3n+3}$,\cite{Lomanova_et_al:2012}
consistent with an earlier extended x-ray absorption fine structure
(EXAFS) analysis of the 5-layer system
Bi$_6$Fe$_2$Ti$_3$O$_{18}$.\cite{MonteroCabrera:2008wu} A preferential
occupation of the outer site with \ti\ cations, has also been reported
for the related system
Bi$_5$CrTi$_3$O$_{15}$.\cite{Giddings_et_al:2011}

It has been argued that the \ti\, cation prefers the more distorted
outer site because of its $d^0$ electron configuration, which exhibits
a strong tendency for off-center displacement due to the second-order
Jahn-Teller effect.\cite{Bersuker:2010} Interestingly, our relaxations
show that, while it is indeed energetically more favorable that Fe
occupies the inner site (at least within PBEsol and at the
experimental volume), the presence of \fe\, on the outer site in fact
creates a significantly more distorted coordination octahedra compared
to the \ti\, cation (see the above discussion on local
tetragonality). This strong local tetragonality, which resembles the
local coordination found in BiFeO$_3$ films under strong compressive
epitaxial strain,\cite{Zeches_et_al:2009, Bea_et_al:2009,
  Hatt/Spaldin/Ederer:2010} suggests that the \fe\ cation also plays
an active role in determining the site preference. The change in site
preference with lattice constants seems to point towards competing
tendencies between the individual site preferences of \fe\ and
\ti\ cations.

Another possible explanation for the outer site preference of \ti,
suggested in Ref.~\onlinecite{Giddings_et_al:2011}, is that it is
electro-statically more favorable to have the higher charged \ti\,
cation located closer to the negatively charged oxygen in the fluorite
layer. Even though the importance of electro-static effects cannot be
strictly ruled out by our results, we point out that the pronounced
dependence on lattice parameters and volume seems to indicate that
changes in chemical bonding and coordination are more relevant than
electro-static effects.

Further insight in the underlying mechanism could be obtained by a
systematic study of site preference as function of volume and strain,
and also by investigating different compositions where \fe\ is replaced
by other magnetic TM cations, such as e.g. Cr$^{3+}$ or Mn$^{3+}$.

\subsection{Internal structural parameters and mode decomposition}
\label{subsec:internal_structure}

Next, we analyze the differences in the internal atomic coordinates of
the various configurations listed in Table~\ref{config}. Due to the
numerous internal degrees of freedom and different symmetries for the
different configurations, we do not compare the individual atomic
coordinates directly. Instead, we perform a symmetry mode
decomposition of all relaxed structures, as well as of the
experimentally observed structure, using the same high symmetry
\highsym\ reference structure in all cases (with equivalent atoms on
all octahedrally coordinated sites). This allows for a more systematic
quantitative comparison of the most relevant degrees of
freedom.\cite{Perez-Mato/Orobengoa/Aroyo:2010}

The position of each atom $\vec{r}_i$ in the unit cell, can be written
as the sum of its position in the high-symmetry reference structure
$\vec{r}_i^0$ (here taken to be the paraelectric \highsym\ structure)
and a displacement vector $\vec{u}_i$:
\begin{equation}
\vec{r}_i = \vec{r}_i^0 + \vec{u}_i \quad .
\end{equation}
The set of displacement vectors can then be expressed as linear
combinations of \emph{mode vectors} $\vec{\epsilon}_{i,m}$ with
amplitudes ${A}_m$:
\begin{equation}
\vec{u}_i = \sum_m A_m \vec{\epsilon}_{i,m} \quad .
\end{equation}
Thereby, each mode vector can be chosen such that it transforms
according to a single irreducible representation, characterized by the
index $m$, of the symmetry group of the reference structure. In
general, these mode vectors consist of several components --- the
symmetry-adapted modes basis for that irreducible representation,
which correspond for example to displacements of cations on different
Wyckoff positions. It is important not to confuse these displacement
modes, which correspond to the distortion in the relaxed structure,
with specific phonon eigenmodes.

In the following, we only discuss the \emph{total} mode amplitude for
each irreducible representation, which is calculated from the
vectorial sum of all the individual components. For more details
please refer to Ref.~\onlinecite{Perez-Mato/Orobengoa/Aroyo:2010}. In
the present work, the mode decomposition is performed using the tool
AMPLIMODES available from the Bilbao Crystallographic
Server.\cite{Orobengoa_et_al:2009}

\begin{figure}
\includegraphics[width=\columnwidth]{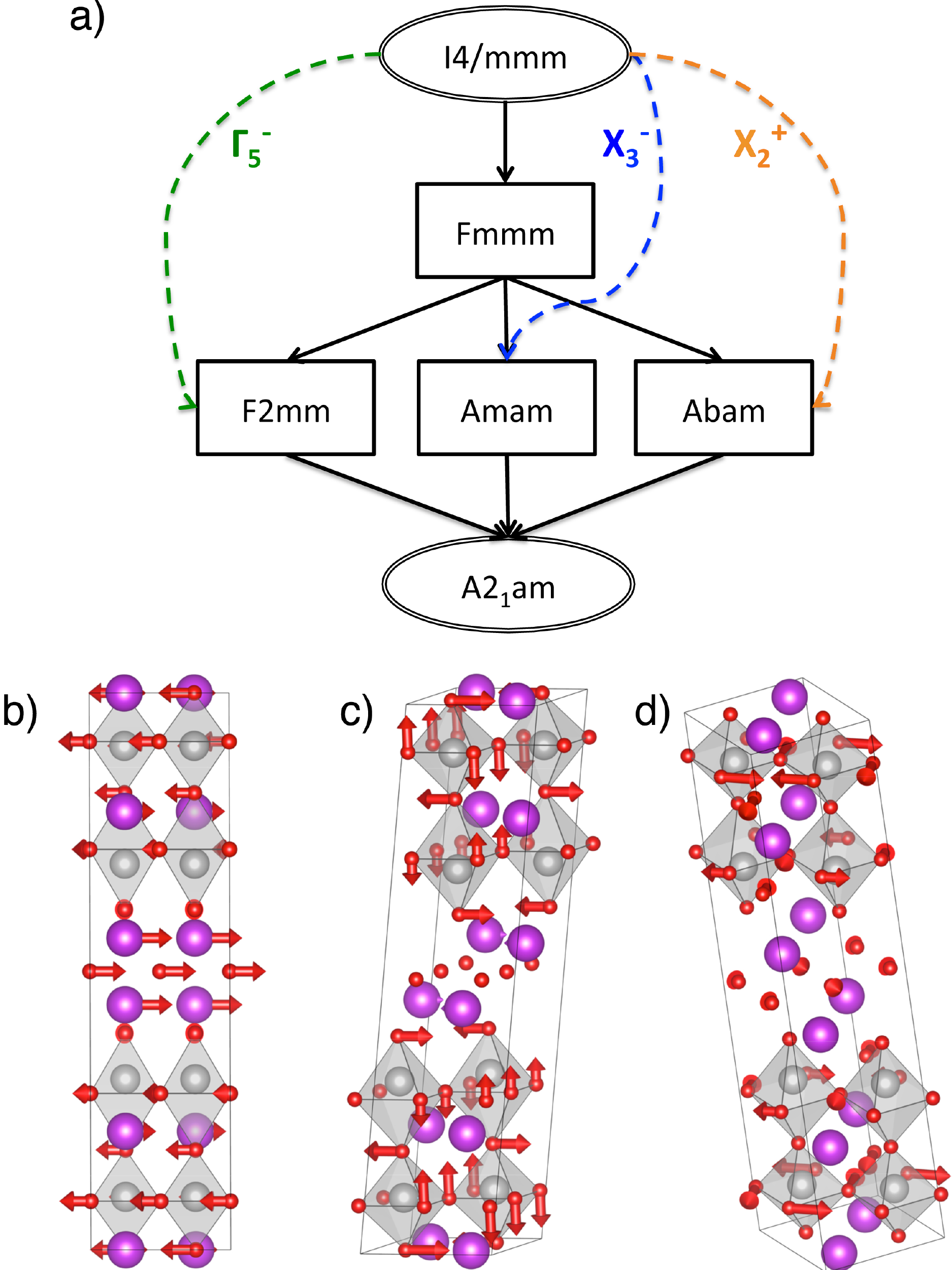}
\caption{(Color online) a) Group-subgroup tree representing all
  possible ways of connecting the paraelectric \highsym\ symmetry to
  the ferroelectric \lowsym\ symmetry. Picture generated with the help
  of SUBGROUPGRAPH.\cite{Ivantchev:2000wk} b-d) Sketches of the three
  main distortion modes: $\Gamma_5^-$, $X_3^-$, and $X_2^+$,
  respectively. The displacement amplitudes of the various ions are
  indicated by the arrows. Different atoms are represented in the same
  way as in Fig.~\ref{tetragonal_Aurivillius}.}
\label{group-tree-and-modes}
\end{figure}

There are three distinct symmetry modes that are involved in the
transition from the paraelectric \highsym\ group to the experimentally
observed ferroelectric \lowsym\ group: the polar $\Gamma_5^-$ mode and
two zone-boundary modes, $X_3^-$ and $X_2^+$ (see
Fig.~\ref{group-tree-and-modes}). This is in direct analogy to the
case of the 2-layered Aurivillius system
SrBi$_2$Ta$_2$O$_9$,\cite{Perez-Mato_et_al:2004} since the high and
low symmetry groups of both cases are identical.  However, as stated
in Sec.~\ref{subsec:method_structure}, the configurations we analyze
in this work exhibit symmetries lower than \lowsym, and therefore
contain additional distortion modes. Nevertheless, as already
mentioned at the end of Sec.~\ref{subsec:method_structure}, we find
that the departure from \lowsym\ symmetry is small. In particular, we
find that for all configurations the three modes listed above are the
most dominant ones, in most cases with significantly larger amplitude
than the ``minor'' modes related to the further symmetry lowering from
\lowsym.

\begin{figure}
\includegraphics[width=\columnwidth]{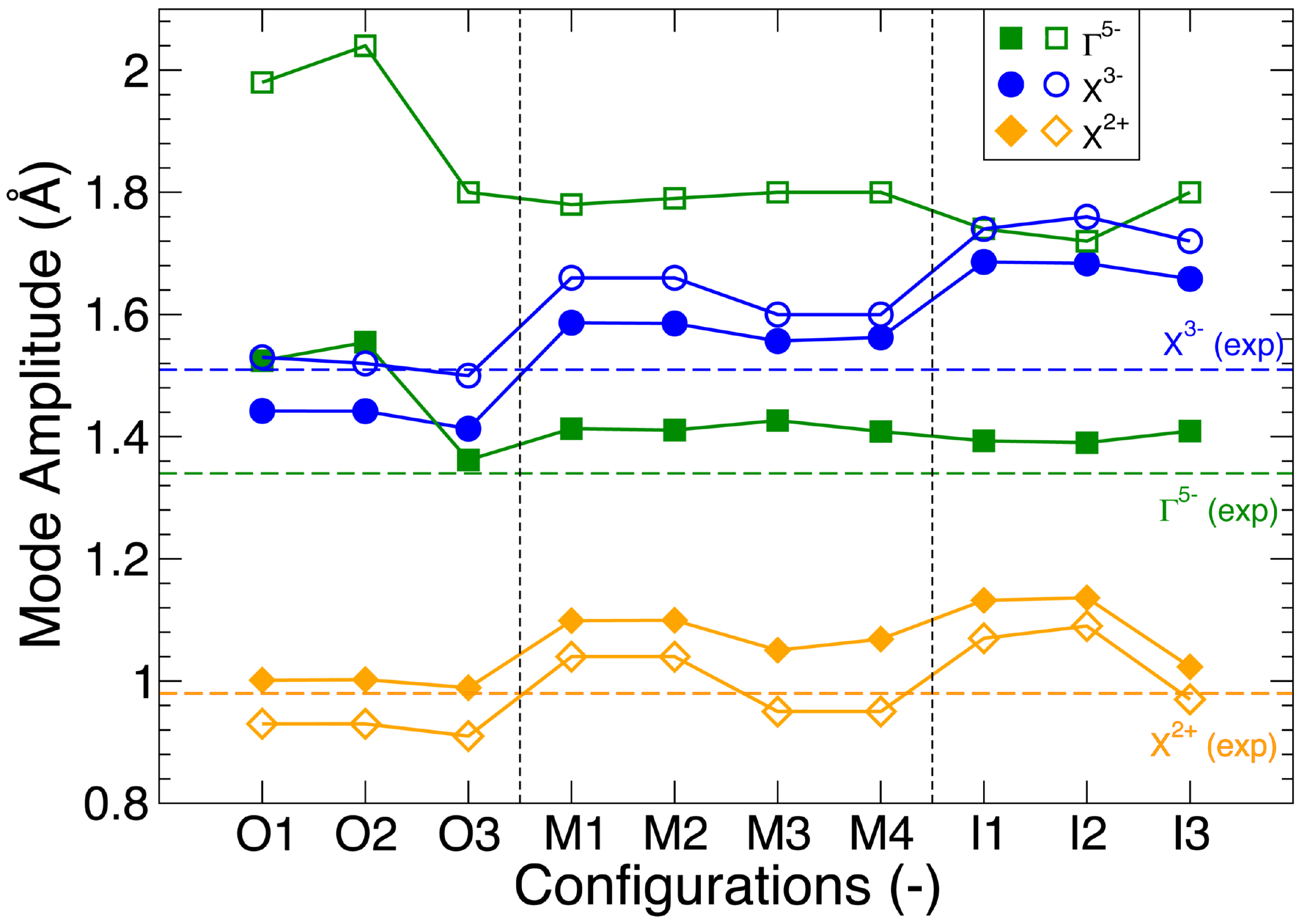}
\caption{(Color online). Amplitudes of the three main distortion modes
  in all considered configurations. Filled and open symbols correspond
  to PBEsol and PBE calculations, respectively. The dashed horizontal
  lines indicate experimental values from
  Ref.~\onlinecite{Hervoches_et_al:2002}. The full lines connecting
  points are simply guides to the eye.}
\label{modeamplitudes}
\end{figure}

Fig.~\ref{group-tree-and-modes}b-d shows the character of the three
modes. These figures correspond to configuration M2, but the main
features are similar for all configurations. It can be seen that the
two zone-boundary modes, $X_3^-$ and $X_2^+$, correspond to ``tilts''
of the oxygen octahedra within the perovskite-like layers around the
in-plane [010] direction and ``rotations'' around the [001] direction,
respectively. Furthermore, the polar $\Gamma_5^-$ mode consists of an
overall displacement of the ions in the \bio\ layer relative to the
ions in the perovskite layer along the in-plane [010] direction. This
resembles the so-called ``rigid layer mode'', which has been
identified as the unstable polar mode in the 2-layer Aurivillius
system
SrBi$_2$Ta$_2$O$_9$.\cite{Stachiotti_et_al:2000,Perez-Mato_et_al:2004}
Similar unstable (or very close to unstable) polar phonons have also
been found in the 1-layer system Bi$_2$WO$_6$
(Refs.~\onlinecite{Machado_et_al:2004,Djani_et_al:2012}) and in the
3-layer system
Bi$_4$Ti$_3$O$_{12}$.\cite{Machado_et_al:2004,Perez-Mato_et_al:2008}
Additional polar instabilities, corresponding to $A$-site or $B$-site
displacive modes within the perovskite layer, have been identified for
Bi$_2$WO$_6$ and Bi$_4$Ti$_4$O$_{12}$,
respectively.\cite{Machado_et_al:2004,Perez-Mato_et_al:2008,Djani_et_al:2012}
We note that the $\Gamma_5^-$ mode depicted in
Fig.~\ref{group-tree-and-modes} also exhibits a strong displacement of
the $B$-site cations in the inner perovskite layer relative to the
center of their coordination octahedra. An analysis of phonon
instabilities in \btfo\, would therefore be instructive to find
further analogies or differences within the Aurivillius
series. However, such an analysis is complicated to perform for \btfo,
due to the lack of a well-defined high-symmetry configuration to be
used in the calculations.

The amplitudes of the three main modes for all configurations are
summarized in Fig.~\ref{modeamplitudes}. It can be seen that even
though there is some variation between the different configurations,
the specific cation distribution does not have too strong an influence
on the relative mode amplitudes. The most striking feature is the
significant overestimation of the $\Gamma_5^-$ mode amplitude within
PBE. This effect can be related to the larger unit cell volume within
PBE, and is thus very similar to the behavior found in many other
ferroelectrics, where it has been shown that the ferroelectric mode is
often very sensitive to the volume and generally becomes more dominant
for larger volumes (see
e.g. Refs.~\onlinecite{Ghosez/Gonze/Michenaud:1996,
  Bhattacharjee/Bousquet/Ghosez:2009, Ederer/Harris/Kovacik:2011}). On
the other hand, there are only small differences between PBE and
PBEsol for the two $X$ modes. Taking into account both lattice
parameters (Fig.~\ref{latt-cst}) and mode amplitudes
(Fig.~\ref{modeamplitudes}), we can conclude that overall PBEsol leads
to good agreement with the experimental structure reported in
Ref.~\onlinecite{Hervoches_et_al:2002}, and is therefore preferable to
PBE for calculating the physical properties of \btfo\, and related
materials.

\subsection{Electric polarization}
\label{subsec:pol}

We now calculate the spontaneous electric polarization of \btfo\,
using the so-called ``Berry-phase'' or ``modern'' theory of
polarization.\cite{King-Smith/Vanderbilt:1993,Vanderbilt/King-Smith:1993,Resta:1994}
Hereafter, the spontaneous polarization is defined as the
difference in polarization between the ferroelectric ground state
structure and a suitably defined centrosymmetric reference. In most
cases, the paraelectric high temperature phase provides a suitable
reference structure for this purpose. However, in the case of \btfo\,
this corresponds to the tetragonal $I4/mmm$ structure shown in
Fig.~\ref{tetragonal_Aurivillius}, which requires a random
distribution of Fe and Ti cations over the different $B$-sites within
each perovskite layer. As discussed in
Sec.~\ref{subsec:method_structure} the specific distributions of Fe
and Ti considered in our calculations lead to symmetries lower than
$I4/mmm$, in some cases even to polar symmetries. In the following we
therefore focus on two representative configurations, I1 and O1,
corresponding to ``inner'' and ``outer'' cases, respectively. Both
configurations correspond to distributions of Fe and Ti cations that
do not break inversion symmetry and thus allow us to define suitable
centrosymmetric reference structures.

These centrosymmetric reference structures are constructed
by removing the polar $\Gamma_5^-$, the $X_2^+$, and all minor
distortion modes from the fully relaxed ferroelectric structures.
For both configurations this leads to non-polar $P2_1/m$ symmetry,
whereas the corresponding fully relaxed structures have polar $P2_1$
symmetry. We note that the spontaneous polarization does not depend on
a specific choice for the reference structure as long as the different
reference structures can be transformed into each other by a
continuous deformation that conserves the insulating character of the
system and does not break inversion symmetry.

Another important consequence of the modern theory of polarization
follows from the arbitrariness in the choice of the unit cell. As a
result, the polarization inherits the periodicity of the crystal and
is only determined modulo a \emph{polarization quantum}, $\Delta P_Q =
\frac{e\vec{R}}{\Omega}$, with $e$ the electronic charge, $\Omega$ the
volume of the unit cell, and $\vec{R}$ the shortest lattice vector
along the polarization direction. Nevertheless, polarization
differences between two structures that are related to each other by a
small deformation are well defined as long as the corresponding change
in polarization is small compared to the polarization quantum. If this
is not the case, then intermediate deformations need to be considered
until each polarization value can be unambiguously assigned to a
specific ``branch'' of the ``polarization lattice'' (see e.g.
Ref.~\onlinecite{Spaldin:2012}). Note that in the present case the
polarization quantum is rather small, $\Delta P_Q \sim
14\,\mu$C/cm$^2$, since the unit cell volume is quite large (due to
the large unit cell length along $\vec{c}$) whereas the polarization
is pointing along the short in-plane direction $\vec{b}$.

\begin{figure}
\includegraphics[width=\columnwidth]{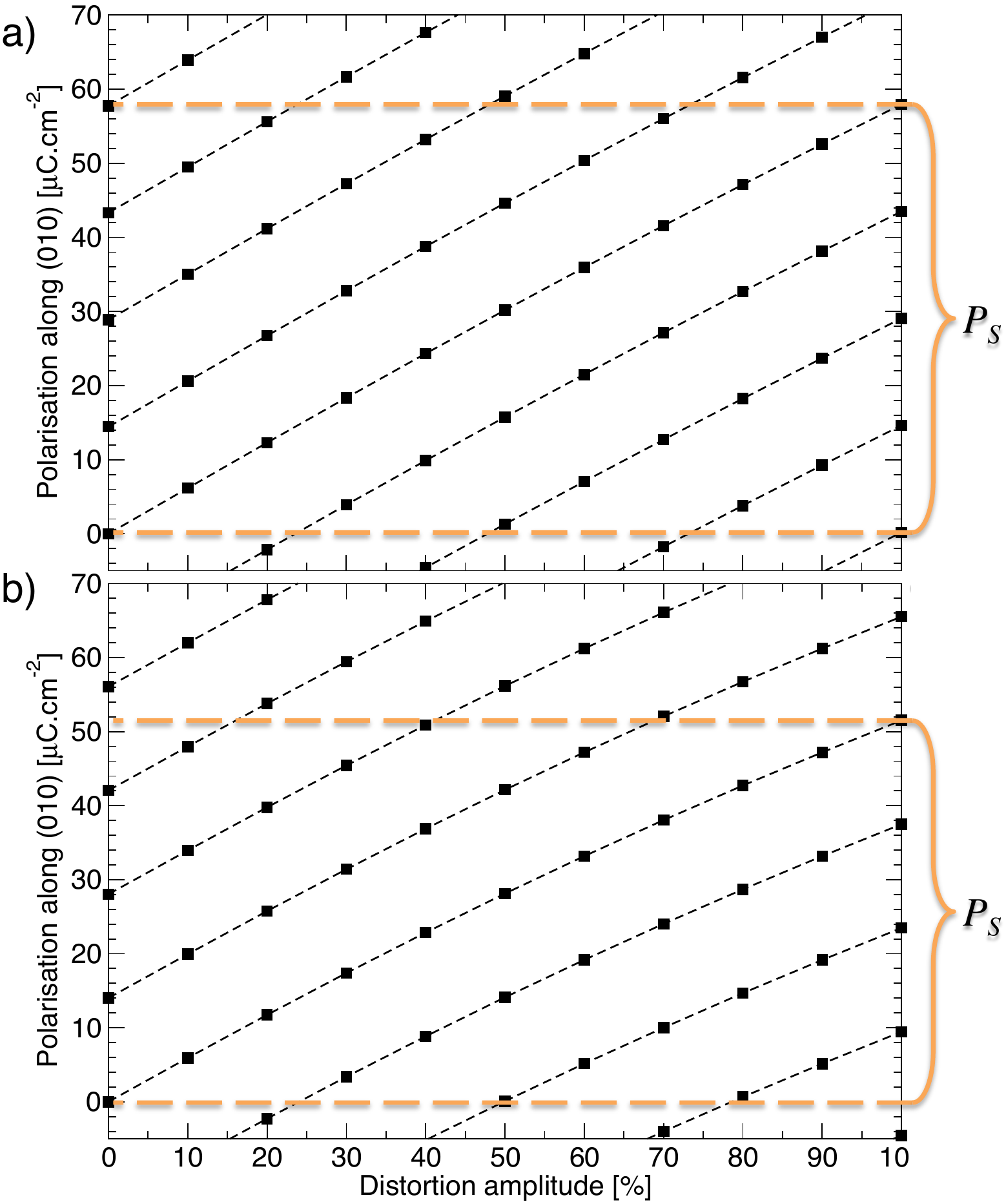}
\caption{Change in polarization for increasing distortion from the
  centrosymmetric $P2_1/m$ structure to the relaxed ferroelectric
  structure with polar $P2_1$ symmetry calculated using PBEsol. Top
  (bottom) panel corresponds to configuration I1 (O1). All branches of
  the polarization lattice within a given range are shown. Points on
  the same branch are connected by lines. The magnitude of the
  spontaneous polarization $P_s$ is also indicated (see main text).}
\label{plot-pol}
\end{figure}

Fig.~\ref{plot-pol} shows the calculated polarization using PBEsol for
the two selected configurations as a function of the distortion
amplitude connecting the centrosymmetric reference (0\%\, distortion
amplitude) with the fully relaxed ferroelectric structure (100\,\%
distortion amplitude). For each distortion amplitude we have included
all values of the polarization lattice within a given range, the
differences being exactly equal to the polarization quantum defined
above. In order to identify the polarization values at different
distortion amplitudes that correspond to the same branch of the
polarization lattice (i.e., how to connect the individual points in
Fig.~\ref{plot-pol} by lines), we evaluate a point charge estimate of
the polarization using the formal charges of the ions (Bi$^{3+}$, \fe,
\ti, O$^{2-}$). While this might give only a rough estimate of the
magnitude of the spontaneous polarization, it clearly shows whether a
positive or negative change in polarization is expected with
increasing distortion. In the present case this is sufficient to make
an unambiguous assignment of branches without the need to consider
further intermediate distortion amplitudes. The resulting branches are
indicated by lines in Fig.~\ref{plot-pol}. The spontaneous
polarization $P_s$ corresponds to the difference between 0 and 100\,\%
distortion evaluated on the same branch. It can be seen that the
polarization quantum (i.e., the distance between two neighboring
branches) is indeed significantly smaller than $P_s$.

\begin{table}
\caption{\label{table-pol} Spontaneous polarization $P_s$ (calculated
  using the Berry phase approach) and point charge estimate $P_{pc}$
  (based on formal ionic charges) together with the amplitude of the
  polar $\Gamma_5^-$ mode for the two representative configurations,
  O1 and I1, calculated using PBE and PBEsol energy functionals.}
\begin{ruledtabular}
\begin{tabular}{llccc}
\multicolumn{2}{c}{}         & $P_s$ & $P_{pc}$ & $\Gamma_5^-$
\\
\multicolumn{2}{c}{}         & $(\rm \mu C/cm^2)$ & $(\rm \mu C/cm^2)$ & (\AA) 
\\ \hline
\multirow{2}{*}{O1} & PBE    & 60.3 & 39.1 & 1.97 
\\
                    & PBEsol & 51.5 & 33.6 & 1.52 
\\ \hline
\multirow{2}{*}{I1} & PBE    & 66.3 & 43.9 & 1.73 
\\
                    & PBEsol & 57.9 & 36.7 & 1.39 
\\ 
\end{tabular}
\end{ruledtabular}
\end{table}

The obtained spontaneous polarizations, $P_s$, together with the point
charge estimates, $P_{pc}$, and the $\Gamma_5^-$ mode amplitudes for
both configurations, calculated using both PBE and PBEsol, are listed
in Table~\ref{table-pol}. In all cases we obtain large values for
$P_s$ between 52 and 66\,$\mu$C/cm$^2$. Due to the overestimation of
the $\Gamma_5^-$ mode amplitude within PBE compared to experiment
(1.34\,\AA), we expect the spontaneous polarization calculated using
the PBEsol functional to be more accurate. We thus predict a
spontaneous polarization of \btfo\, around 55$\mu$C/cm$^2$.

While this is larger than what has been observed so far experimentally
(see e.g. Refs.~\onlinecite{Mao_et_al_JMS:2012} and
\onlinecite{Bai_et_al:2012}), it is quite similar to the spontaneous
polarization for the closely-related 3-layer Aurivillius compound
Bi$_4$Ti$_3$O$_{12}$, as reported both from experiments ($P_S = 50 \rm
\mu C/cm^2$, Ref.~\onlinecite{Cummins:1968ju}) and first-principles
calculations ($P_S = 46 \rm \mu C/cm^2$,
Ref.~\onlinecite{Shah:2010kg}). The difference between our calculated
$P_s \sim 55$\,$\mu$C/cm$^2$ and available experimental data is
therefore likely due to the difficulty in measuring fully saturated
polarization loops in \btfo.

We also observe that the polarization of the I1 configuration is about
10\,\% larger than that of the O1 configuration, for both PBE and
PBEsol. On the other hand, the amplitude of the polar $\Gamma_5^-$
mode is larger for the O1 configuration. This indicates that the
$\Gamma_5^-$ mode is slightly more polar in the I1 configuration than
in the O1 case, i.e. the mode effective charge (see
e.g. Ref.~\onlinecite{Ghosez/Michenaud/Gonze:1998}) of the
$\Gamma_5^-$ mode differs somewhat for the two configurations.

Overall, there is only a moderate influence of the specific Fe
distribution on the ferroelectric properties. Most importantly, the
presence of the magnetic cation is not detrimental to the
ferroelectric polarization, which remains large compared to other
nonmagnetic Aurivillius compounds.

Finally, we note that the large difference between the polarization
calculated using the Berry phase approach and the point charge
estimate based on formal charges indicates a highly anomalous value of
the mode effective charge, which is enhanced by about 50-60\,\%
compared to its formal value. Such enhanced effective charges are
indicative of ferroelectricity that is driven by hybridization
effects.\cite{Ghosez/Michenaud/Gonze:1998}

\subsection{Electronic structure}
\label{subsec:dos}

\begin{figure}
\includegraphics[width=\columnwidth]{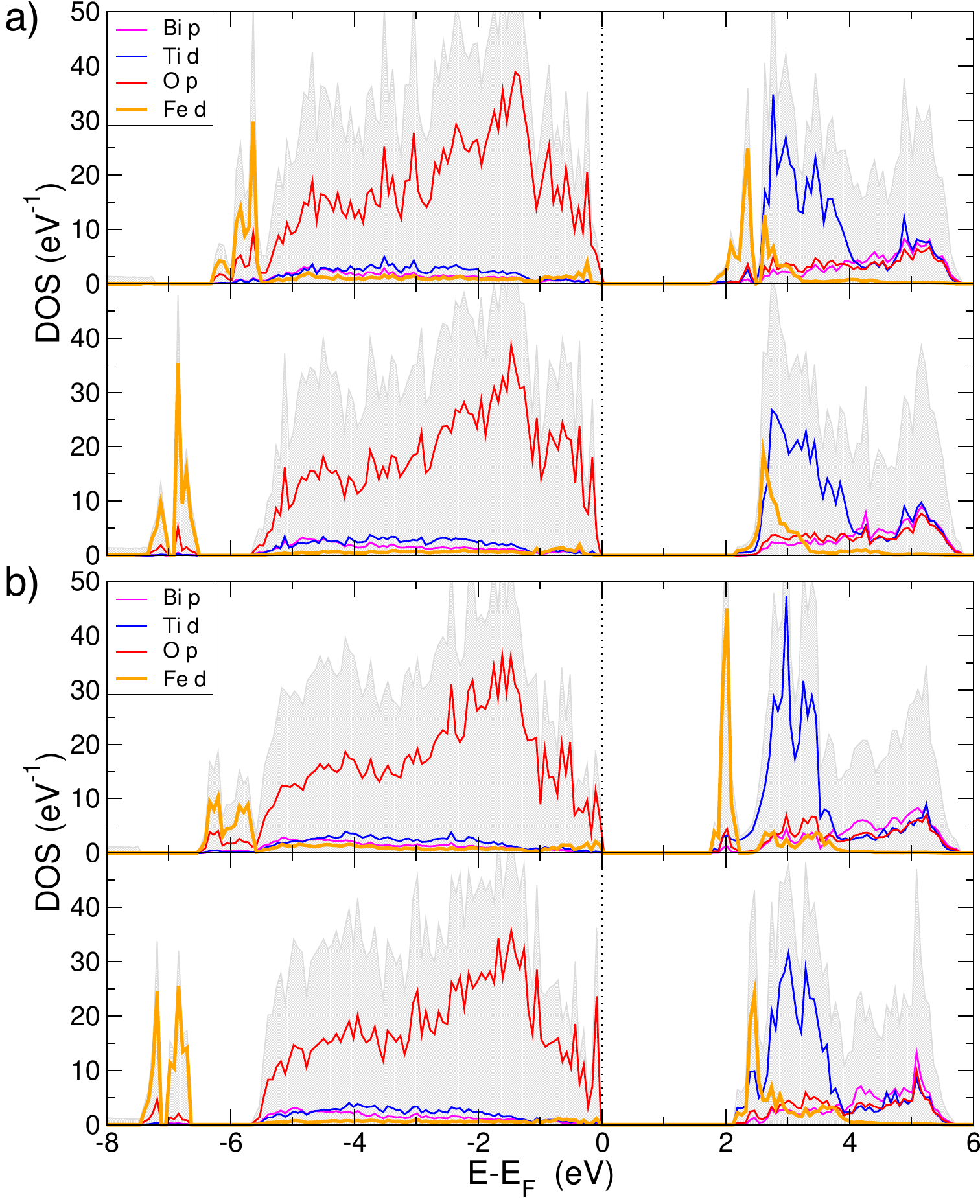}
\caption{(Color online). Total (gray shaded) and projected (lines)
  densities of states (DOS) for configurations O1 (a) and I1 (b) using
  different values of the Hubbard $U$ parameter: $U_{\rm Fe} = 3.0$ eV
  (top panel in each subfigure) and $U_{\rm Fe} = 5.0$ eV (bottom
  panels).}
\label{fig:dos}
\end{figure} 

Fig.~\ref{fig:dos} shows the electronic densities of states (DOS),
total and projected on selected atomic and orbital characters,
calculated using PBEsol for the two representative outer and inner
configurations, O1 and I1. It can be seen that while there are some
differences, the overall features of the densities of states are the
same for the two cases. This indicates that the specific Fe/Ti
distribution has only a marginal impact on the global electronic
structure.

For both configurations we obtain a large bandgap of $1.7$ eV (for
$U_\text{Fe}=3$\,eV). The valence band is formed by states with
predominant O-$p$ character and the conduction band by states with
predominant Ti/Fe-$d$ character (and some Bi-$p$ character at slightly
higher energies). There is a strong splitting ($\sim 7.5$ eV) between
the Fe-$d$ states with different local spin projection, consistent
with a high spin $d^5$ electron configuration of \fe. On the other
hand, the Ti-$d$ states are essentially empty (apart from small
admixtures in the valence band due to hybridization with the O-$p$
states), consistent with a $d^0$ configuration of the \ti\ cation.

In order to test the impact of the Hubbard $U$ parameter on the
electronic structure, we compare the DOS obtained with $U_{\rm Fe} =
3.0$\,eV (the value used throughout this paper) with that obtained for
a somewhat higher value of $U_{\rm Fe} = 5.0$\,eV.  The larger
$U_\text{Fe}$ increases the spin splitting of the Fe $d$ states and
pushes the unoccupied local minority spin Fe $d$ states to higher
energies, so that the lower conduction band edge becomes more
dominated by the Ti $d$ states and the band gap increases to
$2.1$\,eV. Note that an even larger value for $U_\text{Fe}$ will not
lead to a further increase of the bandgap, which is then fixed by the
energy difference between the occupied O $p$ and unoccupied Ti $d$
states. 

Unfortunately, to the best of our knowledge, no spectroscopic data is
available for \btfo, that would allow to identify the orbital
character of the lower conduction band edge. Such data would be
desirable to verify the correct description of the electronic
structure of this material within the DFT+$U$ approach, and to narrow
down the specific value for $U_\text{Fe}$ to be used in such
calculations.

\subsection{Magnetic coupling constants}
\label{subsec:magnetism}

\begin{figure}
\includegraphics[width=\columnwidth]{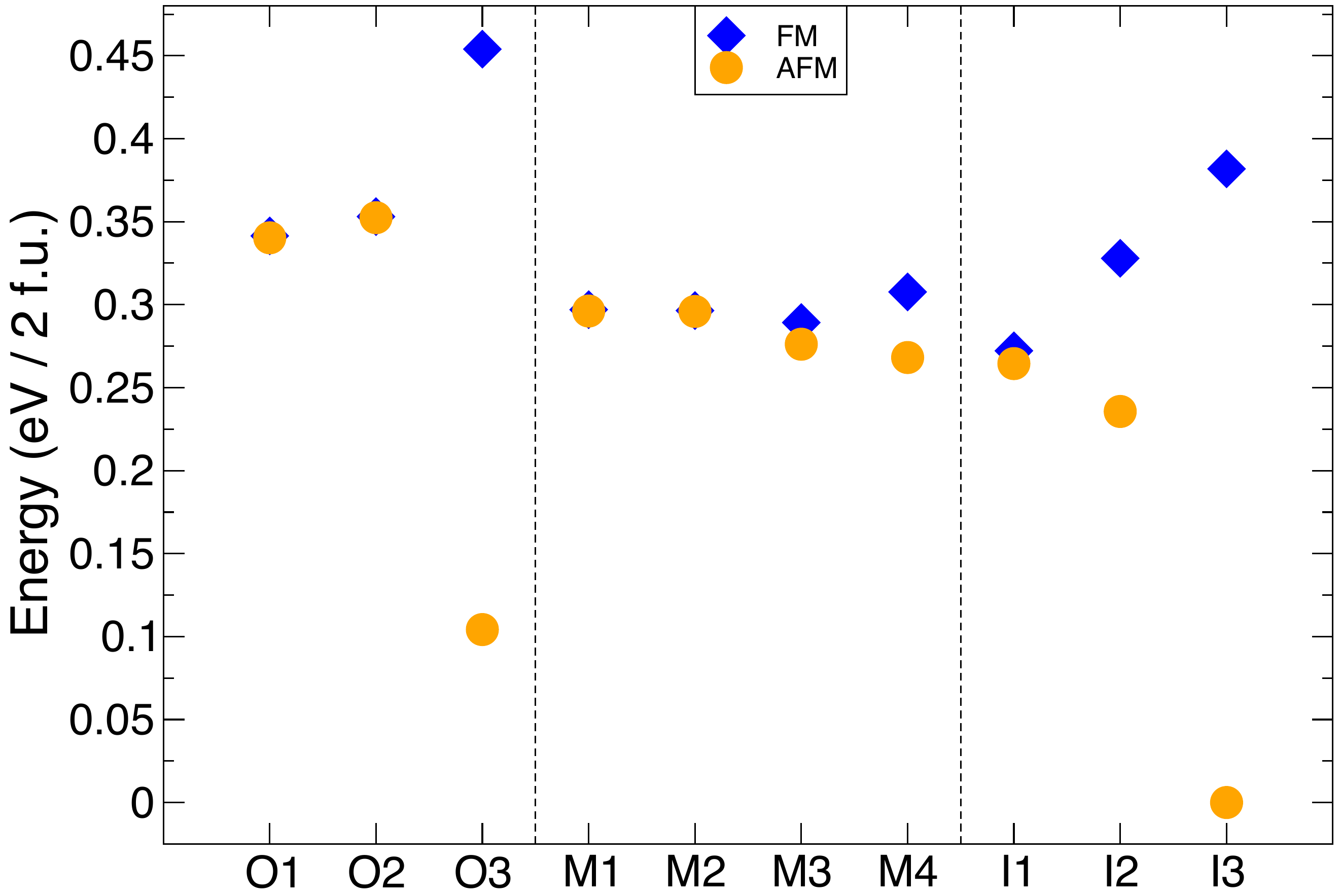}
\caption{(Color online) Energies of all configurations for FM (blue
  squares) and AFM (red circles) orientation of the Fe magnetic
  moments, within PBEsol.}
\label{fig:mag}
\end{figure}

In this section we calculate the strength and character of the
magnetic coupling between the Fe cations in different relative
positions. For this purpose we fully relax all atomic positions and
lattice parameters in each configuration for two different relative
alignments of the magnetic moments of the two \fe\ cations within the
unit cell: i) parallel, i.e., ferromagnetic (FM) alignment, and ii)
antiparallel, i.e., antiferromagnetic (AFM)
alignment. Fig.~\ref{fig:mag} shows the resulting total energies
calculated using PBEsol (PBE leads to very similar results).

It can be seen that for all configurations where the two \fe\, cations
are in nearest-neighbor (NN) positions within the perovskite layer
(O3, M4, I2, and I3), there is a large energy difference between the
FM and the AFM case. For the two configurations where the \fe\,
cations are in next-nearest-neighbor (NNN) positions (M3 and I1),
there is only a small energy difference between FM and AFM, whereas in
all other configurations the energy differences are negligible. In all
cases (even those with essentially negligible energy difference
between FM and AFM) the AFM orientation has lower energy than the FM
orientation.

The strong NN coupling and the rather short range of the magnetic
coupling is indicative of the superexchange mechanism, which is
generally the dominant magnetic coupling mechanism in insulating
oxides such as \btfo. Furthermore, the preferred AFM alignment between
NN cations is consistent with the Goodenough-Kanamori rules for the
\fe\, cation with a $d^5$ electron
configuration.\cite{Anderson:1963,Goodenough:Book}

\begin{table}
\caption{\label{tab:mag} Calculated magnetic Heisenberg coupling
  constants for nearest and next-nearest-neighbor coupling between the
  \fe\ cations. Also listed are the corresponding (average) Fe-O bond
  distances, $d_\text{Fe-O}$, and the deviation, $\Delta \phi$, of the
  Fe-O-Fe bond angle from the ideal value of 180$^\circ$ for the
  nearest-neighbor configurations. Upper part (rows 2-4) corresponds
  to PBE calculations, lower part (rows 5-7) to PBEsol calculations.}
\begin{ruledtabular}
\begin{tabular}{r| c c c c | c c}
configuration & I2 & M4 & I3 & O3 & I1 & M3 \\
\hline
$J$ (meV)      & 43.0 & 11.8 & 42.3 & 39.7 & 0.9 & 1.5\\ 
$d_\text{Fe-O}$ (\AA) & 2.04  & 2.29  & 2.04  & 2.04 & - & -\\ 
$\Delta \phi$ ($^\circ$) & 28.71 & 24.12 & 33.27 & 41.90 & - & - \\
\hline
$J$ (meV) & 46.1 & 19.8 & 47.7 & 43.7 & 1.0 & 1.7 \\
$d_\text{Fe-O}$ (\AA) & 2.03 & 2.20 & 2.00 & 1.99 & - & - \\
$\Delta \phi$ ($^\circ$) & 27.48 & 22.45 & 32.74 & 41.53 & - & -
\end{tabular}
\end{ruledtabular}
\end{table}

For a more quantitative analysis of the strength of the magnetic
coupling, we extract coupling constants $J_{ij}$ by mapping the
calculated energy differences onto a Heisenberg model, $E =
\sum_{\left<ij\right>} J_{ij} \hat{e}_i \hat{e}_j$, where the sum is
over all ``bonds'' and the normalized vector $\hat{e}_i$ indicates the
direction of the magnetic moment of the magnetic ion with index
$i$. For this purpose, both FM and AFM energies are calculated using
the relaxed structures obtained with AFM alignment of magnetic
moments, in order to remove the structural component from the
determination of the magnetic coupling constants. Since we consider
only coupling between Fe in NN and NNN positions, each coupling
constant is associated with exactly one particular ionic
configuration. The resulting coupling constants are summarized in
Table~\ref{tab:mag}.

We obtain rather strong nearest-neighbor couplings of the order of
$J_\text{NN} \sim$ 40-50\,meV (except for configuration M4) and
relatively weak next-nearest-neighbor coupling of $J_\text{NNN} \sim$
1-2\,meV. The couplings are slightly stronger in PBEsol compared to
PBE, due to the slightly smaller bond distances within
PBEsol. However, the differences between the two functionals are not
very large. The nearest-neighbor coupling between inner and outer
perovskite layers (configuration M4) is smaller than the other three
nearest-neighbor couplings due to the strong sensitivity of the
superexchange interaction to the bond distances and bond angles
connecting the two coupled magnetic cations. The Fe-O bond distances,
averaged over the two Fe-O bonds connecting two nearest-neighbor Fe
cations ($d_\text{Fe-O}$), and the corresponding Fe-O-Fe bond angles
($\Delta \phi$) are also listed in Table~\ref{tab:mag}. It can be
seen, that the average Fe-O bond-length between adjacent inner and
outer Fe sites is significantly larger for configuration M4 than for
the other nearest-neighbor configurations. This is related to the
strong elongation of the oxygen octahedra surrounding the outer Fe
site.

The calculated nearest-neighbor coupling between the Fe cations in
\btfo\, is comparable in both sign and magnitude to the coupling in
BiFeO$_3$, $J = 50.3$\,meV, which we calculate using PBEsol and
$U_\text{Fe}=3$\,eV (compare also to
Ref.~\onlinecite{Baettig/Ederer/Spaldin:2005}). We also evaluate the
impact of the Hubbard $U$ parameter on the strength of the magnetic
coupling by recalculating the coupling constants for $U_{\rm Fe} =
5.0$\, eV using PBE. This has the effect of reducing the average $J$
by $\sim 30 \%$. For example, the magnetic coupling corresponding to
configuration I2 is $J = 43.0$\, eV with $U_\text{Fe}=3$\,eV, and
becomes $J = 31.2$\, eV using $U_\text{Fe}=5$\,eV. Such a decrease of
the magnetic coupling strength with $U_\text{Fe}$ is expected with
superexchange as the dominant coupling mechanism. Within the simplest
model of superexchange, the coupling strength is proportional to
$t^2/U$,\cite{Anderson:1963} where $t$ is the effective hopping
between Fe-$d$-like Wannier states and $U$ is the corresponding
Hubbard interaction parameter. While there is no direct correspondence
between this $U$ and $U_\text{Fe}$ used in the DFT+$U$ calculations,
this simple picture is sufficient to explain the overall decrease of
the magnetic coupling constants with $U_\text{Fe}$.

Due to the relatively low concentration of magnetic cations in \btfo\,
(25\,\% on the octahedrally coordinated perovskite-like sites) and the
short range of the magnetic superexchange interaction, it is unclear
whether magnetic long range order can occur in this system. One
requirement for long range order is that percolation can be achieved
throughout the material, i.e. the majority of Fe cations need to be
connected with each other through continuous links between coupled
neighbors. In a three-dimensional simple cubic lattice, to which the
4-perovskite thick layers can be related, the threshold for
percolation is 31\% of magnetic sites when considering only NN
couplings, and 14\% with both NN and NNN
couplings.\cite{Kurzawski/Malarz:2012} Thus, with 25\,\% of magnetic
sites as in \btfo, percolation is dependent on the rather weak NNN
coupling, which will probably give rise to magnetic order only at
rather low temperatures. Furthermore, the perovskite-like layers are
separated by Bi$_2$O$_2$ layers that do not contain any magnetically
active ions. The shortest superexchange path connecting two Fe cations
on opposite sides of the Bi$_2$O$_2$ layer is of the form Fe-O-O-O-Fe,
containing two O-O links, which will only give rise to a very weak
coupling.

We can obtain a first estimate of the strength of this coupling from
the energy difference between FM and AFM orientations for
configuration O1. This energy difference is directly related to the
interlayer coupling if one assumes that the coupling of the two
\fe\ cations through the perovskite layer is negligible. This seems
reasonable, since the corresponding superexchange path involves a
total of 8 TM-O bonds. We obtain $J_{\rm interlayer} = 0.6$\,eV, which
is small but comparable with the weak next-nearest-neighbor
coupling. However, we note that further calculations with a doubled
unit cell containing 96 atoms are necessary to double-check this
result and obtain a more accurate estimate of the interlayer coupling.


\section{Summary and conclusions}
\label{sec:summary}

In summary, we have calculated from first-principles the basic
stuctural, ferroelectric, and magnetic properties of the Aurivillius
phase \btfo. We have also addressed a potential site preference of the
\fe\ and \ti\ cations within the perovskite-like layers and discussed
the possibility for magnetic long-range order within this material.

After systematically comparing our results obtained within PBE and
PBEsol with available experimental data, we conclude that the PBEsol
exchange-correlation functional provides an excellent description of
the structural degrees of freedom in \btfo, superior to PBE, and we
suggest to use PBEsol for further DFT studies of related members of
the Aurivillius family.

Our results show that there is a preference for the \fe\ cations to
occupy the inner sites within the perovskite-like layers, consistent
with recent M\"ossbauer data.\cite{Lomanova_et_al:2012} We have also
shown that this site preference depends strongly on the lattice
constants and reverses to an outer site preference for the slightly
more extended lattice parameters within PBE. In addition, the $c/a$
ratio of the crystallographic unit cell is systematically larger for
configurations where \fe\ occupies the outer sites. This suggests the
possibility to control the site occupancies via epitaxial strain in
thin films of \btfo.

Moreover, we calculate a large value of 55\,$\mu$C/cm$^2$ for the
spontaneous electric polarization, which shows that the presence of
the nominally non-ferroelectric magnetic \fe\ cation does not impede
the ferroelectricity in this Aurivillius system.

Finally, we show that there is a strong antiferromagnetic coupling
between \fe\ cations in nearest-neighbor positions, characteristic of
the superexchange interaction between $d^5$ cations, but that the
coupling between further neighbors is rather weak and essentially
becomes negligible beyond second neighbors. With a concentration of
25\,\% magnetic cations on the perovskite $B$-site, the system is
above the percolation threshold for a simple cubic lattice with both
NN and NNN interactions, but it is unclear how the interlayer
\bio\ layers will influence percolation in the Aurivillius structure
and what ordering temperature can be expected based on the weak NNN
coupling. In addition, it is likely that the percolation within the
perovskite layers depends critically on the distribution of magnetic
cations within these layers. The possibility to control this
distribution either by strain or by changing preparation conditions,
as discussed in Sec.~\ref{subsec:global_structure}, could therefore be
crucial for achieving magnetic long range order in \btfo\ and related
Aurivillus systems.

We note that even with a relatively weak interlayer coupling, long
range order at elevated temperatures can in principle be achieved if
the magnetic moments within the perovskite layers are highly
correlated (due to strong magnetic coupling \emph{within} these
layers). In this case, all weak interlayer couplings will add up
constructively and lead to a relatively strong effective coupling
between the layers. Such behavior has been studied for example in the
context of the anisotropic quasi-two-dimensional Heisenberg model (see
e.g. Ref.~\onlinecite{Yasuda:2005dw}).

It therefore seems that, in order to achieve robust magnetic order
above room temperature, a higher concentration of magnetic ions would
be desirable, which would lead to stronger coupling within the
perovskite layers. One obstacle for achieving such higher
concentrations of magnetic cations is the relation between the average
valence of the $B$-site cations and the number of perovskite layers
$m$ throughout the Aurivillius series. With a formal charge of $3+$ on
the $A$-site (corresponding to Bi$^{3+}$) the required average valence
on the $B$-site varies from $4+$ for $m=3$ to $3+$ for $m \rightarrow
\infty$.
The possibility to substitute Mn$^{4+}$ for \ti\ in the 3-layer
structure has been studied computationally by Tinte and
Stachiotti.~\cite{Tinte:2012dj} However, they found only incipient
ferroelectricity in the fully substituted system $\rm
Bi_4Mn_3O_{12}$. Furthermore, the Mn$^{4+}$ cation has a rather small
ionic radius, which falls outside the range that is considered
suitable for incorporation on the $B$-site of the Aurivillius
structure.\cite{Newnham/Wolfe/Dorrian:1971}
Magnetic cations that are suitable for incorporation on the $B$-site
within the Aurivillius structure generally have charge states of only
$3+$ or even $2+$. One possible route to increase the concentration of
magnetic cations is therefore to focus on compositions with a higher
number of perovskite layers, i.e. $m > 4$. Indeed, magnetic long range
order has been reported for several $m=5$
systems.\cite{Jartych_et_al:2013,Keeney_et_al:2013} Another
possibility to increasing the \fe\ content (or more generally the
content of magnetic $3+$ cations) in the $m = 4$ layer system would be
to simultaneously replace \ti\ by other $B$-site cations with higher
valence, such as Ta$^{5+}$, Nb$^{5+}$, or W$^{6+}$, leading to
compositions such as Bi$_5$Fe$_{1+x}$Ti$_{3-2x}$Nb$_x$O$_{15}$.

We conclude, that the Aurivillius family is a promising class of
compounds to search for new multiferroics with good ferroelectric and
magnetic properties. The main challenge is to increase the content of
magnetic cations, in order to obtain robust magnetic long range order
at elevated temperatures. The Aurivillius family provides enough
chemical flexibility to explore new compositions that are promising in
that respect. Our results indicate that it is conceivable to have
ferroelectricity coexist with long-range magnetic order, and we hope
that our work will stimulate further research on Aurivillius compounds
as potential room-temperature multiferroics.

\begin{acknowledgments}
This work was supported by the Swiss National Science Foundation under
project no. 200021\_141357, by ETH Z\"urich, and by Science Foundation
Ireland through the FORME project. We thank Lynette Keeney, Roger
Whatmore, and Martyn Pemble from the Tyndall National Institute in
Cork, Ireland, for many important discussions and for pointing our
attention to the Aurivillius phases as potential multiferroics. We
also thank Eric Bousquet for many useful discussions.
\end{acknowledgments}

\bibliography{references}

\end{document}